# The GRW flash theory: a relativistic quantum ontology of matter in space-time?

Michael Esfeld and Nicolas Gisin

Department of Philosophy, University of Lausanne

Michael-Andreas.Esfeld@unil.ch

Group of Applied Physics, University of Geneva

Nicolas.Gisin@unige.ch

(draft 7 October 2013)[1]

**Abstract**

John Bell proposed an ontology for the GRW modification of quantum mechanics in terms of flashes occurring at space-time points. The paper spells out the motivation for this ontology, enquires into the status of the wave-function in it, critically examines the claim of its being Lorentz-invariant and considers whether it is a parsimonious, but nevertheless physically adequate ontology.

*Keywords*: Bell's theorem, Bohmian mechanics, collapse of the wave-function, dispositionalism, GRW flash, GRW mass density, Humeanism, Lorentz-invariance, primitive ontology, propensities, relativistic collapse theory

## 1. Introduction

The modification of the Schrödinger dynamics set out by Ghirardi, Rimini and Weber (GRW) (1986) is the only precisely elaborated proposal for a quantum theory that postulates a dynamics of what is known as the collapse of the wave-function in order to account for measurement outcomes (forerunners of this proposal include Gisin 1984, further developments notably Ghirardi, Pearle and Rimini 1990). Bell (1987, reprinted in 2004, ch. 22) took up this proposal and developed an ontology for it, which is known today as the GRW flash ontology (GRWf). The term "flash", however, is not Bell's, but was introduced by Tumulka (2006). This ontology has recently met with considerable interest.[2] The main reasons for this interest are the following two ones: (a) GRWf is a clear and parsimonious proposal for a quantum ontology that admits only one actual distribution of matter in space-time. (b) It seems that Bell's expectation that GRWf can be turned into a fully relativistic theory did materialize.

The aim of this paper is to assess this ontology, notably the claim of its being fully relativistic. In the next section, we will set out the primitive ontology of flashes and enquire into the status of the wave-function (section 2). We will then examine the formulation of a relativistic GRWf theory by Tumulka (2006, 2007, 2009, section 4) and maintain that this proposal achieves the aim of a Lorentz-invariant quantum theory of matter in space-time only if one limits oneself to considering possible entire distributions of flashes, renouncing an account of the coming into being of the actual distribution of the flashes (section 3). Against this background, we will argue that a fully relativistic ontology of GRWf is only possible if one endorses Humeanism about laws of nature (section 4). Finally, we will show that even before it comes to including a physical account of interactions in the GRWf theory, there is a problem how to conceive interactions at all in this theory. Our conclusion therefore is that the parsimony of this ontology is not an advantage, but rather a serious obstacle for its being a convincing proposal for an ontology of quantum physics (section 5).

---

[1] We are grateful to Jeff Barrett, Matthias Egg, Roderich Tumulka, Christian Wüthrich and the participants of the summer school "Physics and philosophy of time" in Saig (Germany) in July 2013 for discussion.

[2] See e.g. Allori et al. (2008), Conway and Kochen (2006, 2009), Maudlin (2011, ch. 10), Wüthrich (2011, pp. 387-389), Arntzenius (2012, ch. 3.15), Placek (2013).

2.     *Flashes as the primitive ontology and the status of the wave-function*

GRWf falls within what is known as primitive ontology approaches to quantum mechanics (QM):[3] an ontology of matter distributed in three-dimensional space or four-dimensional space-time is admitted as the referent of the QM formalism, and a law for the temporal development of the distribution of matter is formulated. The primitive ontology cannot be inferred from the formalism of textbook QM (or the formalism of the GRW theory), but has to be put in as the referent of that formalism. The motivation for doing so is to obtain an ontology that can account for the existence of measurement outcomes – and, in general, the existence of the macroscopic objects with which we are familiar before doing science. Hence, what is introduced as the primitive ontology in space-time has to be such that it can constitute measurement outcomes and localized macroscopic objects in general. That is why the primitive ontology consists in one actual distribution of matter in space at any time, and the elements of the primitive ontology are localized in space-time, being "local beables" in the sense of Bell (2004, ch. 7), that is, something that has a precise localization in physical space at a given time.

Bohm's quantum theory, going back to Bohm (1952) and known today as Bohmian mechanics (BM) (see the papers in Dürr, Goldstein and Zanghì 2013a) is the oldest and most widely known primitive ontology approach to QM. BM endorses particles as the primitive ontology, adding the position of the particles as additional, so-called hidden variable to the formalism of standard QM. Consequently, there is at any time one actual configuration of particles in three-dimensional space, and the particles move on continuous trajectories in physical space. BM therefore needs two laws: the guiding equation fixing the temporal development of the position of the particles, and the Schrödinger equation determining the temporal development of the wave-function. These two laws are linked in the following manner: the role of the wave-function is to fix the velocity of each particle at any time *t* given the position of all the particles at *t*. The theory is non-local, since, due to the entanglement of the wave-function, the velocity of *any* particle at *t* depends, strictly speaking, on the position of *all* the other particles at *t*.

By contrast to BM, the GRW quantum theory modifies the Schrödinger equation (or the Dirac equation) such that this law includes what is known as the collapse of the wave-function. This law then is linked to the primitive ontology in the following manner in GRWf: whenever there is a spontaneous localization of the wave-function in configuration space, that development of the wave-function in configuration space represents an event occurring in physical space, namely there being a flash centred around a space-time point. The flashes are all there is in space-time. That is to say, apart from when it spontaneously localizes, the temporal development of the wave-function in configuration space does not represent the distribution of matter in physical space. It represents the objective probabilities for the occurrence of further flashes, given an initial configuration of flashes. As in BM, there hence are no superpositions of anything existing in physical space. However, by contrast to BM, GRWf does not admit a continuous distribution of matter: there are only flashes being sparsely distributed in space-time, but no trajectories or worldlines of anything.

---

[3]     This term goes back to Dürr, Goldstein and Zanghì (1992, reprinted in Dürr, Goldstein and Zanghì 2013, ch. 2, see end of section 2.2). See Allori et al. (2008) for details.

This is also the main difference between GRWf and the other proposal for an ontology of the GRW theory, namely the ontology of a continuous mass density distribution (GRWm) set out by Ghirardi, Grassi and Benatti (1995) (see also Monton 2004): on GRWm, the wave-function in configuration space and its temporal development according to the GRW equation represent at any time the density of matter (mass) in physical space. The spontaneous localization of the wave-function in configuration space represents a spontaneous contraction of the mass density in physical space, thus accounting for measurement outcomes and well localized macroscopic objects in general. Although GRWf and GRWm are rival proposals for an ontology of the same formalism (the GRW quantum theory), there also is a difference between them on the level of the formalism: if one endorses the GRWm ontology, it is reasonable to pursue a formalism of a *continuous* spontaneous localization of the wave-function (as done in Ghirardi, Pearle and Rimini 1990), whereas if one subscribes to the GRWf ontology, there is no point in doing so (cf. Maudlin 2007b, p. 3167).

In GRWf, as in Bell's version of Bohm's theory, the only local beable is the position of the primitive stuff in space-time (see Bell 2004, ch. 4, and Norsen 2013). Measurements of all other observables have to be accounted for in terms of the position of the primitive stuff. Thus, on GRWf, considering the EPR experiment with a pair of spin 1/2 particles in the singlet state, the only difference between the outcome "spin up" and the outcome "spin down" consists in the flash accounting for the outcome "spin up" occurring in a different position than a flash accounting for the outcome "spin down" in the same wing of the experiment.

What is the status of the wave-function in GRWf? It cannot be regarded as an entity that exists in space-time in addition to the primitive ontology, that is, the flashes. It is not possible to conceive it as a field in physical space, since it cannot be considered as assigning values to space-time points. The wave-function is a mathematical entity defined on configuration space by contrast to a physical entity existing in space-time. Its role in GRWf is the following one: given the specification of an initial configuration of flashes, attributing a wave-function to that configuration and putting that wave-function into the GRW equation makes it possible to calculate probabilities for the further development of the distribution of the flashes in space-time. In other words, given an initial configuration of flashes, the wave-function enables conceiving histories of the distribution of flashes in space-time and makes it possible to assign probabilities to these histories. Note, however, that on the GRWf theory, only one such history actually occurs, and that the initial configuration of flashes does not determine a unique wave-function; any initial configuration of flashes is compatible with many different wave-functions.

The role of the wave-function thus is a nomological one: it is part and parcel of the law that describes the temporal development of the distribution of the flashes in physical space. Its job is to yield the objective, mind-independent probabilities for the occurrence of further flashes, given an initial configuration of flashes and given that the wave-function attributed to that configuration is fed into the GRW law. The same holds for the universal wave-function in BM, as argued by Dürr, Goldstein and Zanghì (2013a, ch. 12): given an initial particle configuration, the job of the wave-function is to yield the velocity of the particles, if fed into the Bohmian guiding equation.

Of course, simply granting a nomological status to the wave-function leaves open all the questions about how laws of nature are grounded in physical reality. As regards the GRW law, one can adopt any of the three main stances that are discussed in the metaphysics of laws. Since the differences between these stances will become relevant when considering whether the GRWf theory can be made Lorentz-invariant, let us briefly outline them as applied to GRWf:

- *Humeanism*: There only is the mosaic of flashes in space-time. Everything else supervenes on that mosaic. Given an initial configuration of flashes, there is nothing in nature that determines the temporal development of the distribution of flashes in space-time, or that, more generally speaking, poses constraints on how that distribution can develop. The flashes simply are there. Considering the distribution of flashes in space-time as a whole, that distribution may exhibit certain patterns or regularities. Based on such contingent regularities, the GRW law with the universal wave-function figuring in it supervenes on those patterns or regularities, being the description of the entire distribution of flashes in space-time that achieves the best balance between simplicity and empirical content.[4]

- *Dispositionalism*: Over and above the flashes being the primitive stuff in physical space, the initial configuration of flashes instantiates a dispositional property – more precisely a propensity – that fixes probabilities for the occurrence of further flashes. The occurrence of such further flashes is the manifestation of that propensity. The propensity of any given configuration of flashes to manifest itself in the occurrence of further flashes is represented by the wave-function. The GRW law supervenes on that propensity in the sense that whenever such a propensity is instantiated in a possible world, the GRW law holds in that world. By contrast to what is admitted by Humeanism, that disposition or propensity hence is a modal property.[5]

- *Primitivism*: Laws are primitive in that they neither reduce to the best description of the distribution of the local matters of particular fact (the Humean mosaic) in a world, nor are they grounded in dispositional properties or propensities instantiated in a world. On primitivism about laws, over and above an initial configuration of flashes, each possible world instantiates the property that a certain law holds in this world, with the wave-function figuring in that law. Given the fact that a certain law is instantiated and the initial configuration of flashes, the probabilities for further occurrences of flashes follow (as on dispositionalism and the propensity view) (see notably Maudlin 2007a for primitivism about laws).

---

[4] See Frigg and Hoefer (2007) for a Humean view of the GRW probabilities. See Esfeld et al. (2013, section 3), Miller (2013) as well as Callender (unpublished, section 5) for applying Humeanism to primitive ontology approaches to QM, notably BM.

[5] See Dorato and Esfeld (2010) for setting out the view of the GRW probabilities being grounded in propensities. See Placek (2013) for further developing that view and Suárez (2007) for propensities in QM in general. See furthermore Belot (2012, pp. 77-80) and Esfeld et al. (2013, sections 4-5) for applying dispositionalism to primitive ontology approaches to QM.

## 3. *A Lorentz-invariant quantum ontology of matter distributed in space-time?*

A central open issue is how primitive ontology approaches to QM fare with respect to relativistic space-time. For present purposes, it is sufficient to consider the flat space-time of special relativity and to demand Lorentz-invariance in the following sense: there is no globally preferred reference frame and thus no privileged foliation of space-time into three-dimensional spatial hypersurfaces that are ordered in one-dimensional time. Hence, there is no absolute simultaneity. The problem that relativistic space-time poses for primitive ontology approaches to QM is brought out by Bell's theorem (1964, reprinted in Bell 2004, ch. 2): in brief, this theorem establishes that in certain situations – such as the EPR experiment –, the probabilities for what happens at a given space-time point are not entirely fixed by what there is in the past light-cone of that point, but depend also on what measurements take place at a spacelike separated interval and what the outcomes of these measurements are (see Bell 2004, ch. 24, and Seevinck and Uffink 2011 for precisions). Bell's theorem is therefore widely taken to show that in QM, events that are separated from a given space-time point by a spacelike interval can contribute to determining what happens at that point. Maudlin (2011, chs. 1-6) analyses this determination in terms of non-local influences (see in particular pp. 118-119, 135-141). Note that the issue of how QM fares with respect to relativistic space-time arises only once one has specified what one takes the primitive ontology of QM to be. If one does not do so, one cannot even formulate the question of whether or not a Lorentz-invariant quantum ontology of matter in space-time can be achieved.

All the worked out primitive ontology approaches to QM meet the operational requirements of special relativity, that is, they do not allow for superluminal signalling. However, the impossibility of superluminal signalling does not show that the ontology is Lorentz-invariant. A theory may exclude superluminal signalling and yet be committed to absolute simultaneity. Superluminal signalling may simply be ruled out due to restrictions on the information that a local observer can obtain. Consider BM and GRWm: if a local observer had complete knowledge of the behaviour of matter in the space-time region where she is situated – the particle trajectories on BM, or the development of the shape of the mass density on GRWm (e.g. its sudden local disappearance due to a collapse elsewhere in space) –, she could infer from that local knowledge what is going on at spacelike distances *at the same time* (see Maudlin 2007b, pp. 3167-3168, and 2008, pp. 161-170). But these theories exclude that a local observer can acquire such knowledge.

By contrast, consider GRWf: even if a local observer had complete knowledge of the occurrences of the flashes in the space-time region where she is situated, that knowledge would never make inferences possible about what there is at spacelike distances. In other words, as Bell already proved at the end of the paper in which he introduced what is known today as the GRW flash ontology (1987), GRWf meets the requirement of relative time translation invariance (Bell 2004, pp. 206-209; see also Maudlin 2011, pp. 239-243). The reason is that according to the GRWf ontology, there is no continuous distribution of matter in space-time – no trajectories or worldlines of particles, no mass density field –, but only flashes occasionally occurring at space-time points. Hence, if one searches for a Lorentz-invariant quantum ontology of matter in space-time, GRWf offers a better prospect of achieving that aim than does BM or GRWm (but see Bedingham et al. 2013 and Dürr et al. 2013b for recent attempts at developing relativistic versions of GRWm and BM).

Indeed, Tumulka (2006 and 2009, section 4) has set out a relativistic version of the GRW flash ontology (rGRWf): he considers an initial configuration or set of seed flashes and an initial wave-function attributed to that configuration (formulated with respect to an arbitrary hypersurface). He puts that wave-function into the GRW equation (more precisely the GRW amendment of the Dirac equation), thus obtaining entire distributions or histories of flashes in space-time with a certain probability attached to each such distribution or history. He shows that any such distribution or history as a whole can be represented in a Lorentz-invariant manner, because the rGRWf law is Lorentz-invariant. The theory does not include interactions as it stands.

Tumulka's rGRWf theory is in a certain sense not a collapse theory: the collapse of the wave-function is not part of the ontology of this theory. Only an initial configuration of flashes and the initial wave-function as figuring in the rGRWf law are necessary to obtain histories of flashes in space-time and probabilities attached to them. By contrast to an Everett-type theory, according to GRWf, exactly one such history of flashes in space-time is real, constituting the actual distribution of matter. Nonetheless, wave-function collapse is not necessary to obtain an ontology of one actual distribution of flashes: GRWf, like any other primitive ontology approach to QM, starts from the assumption that there is exactly one distribution of matter in space-time. There is no question here of an ontology that admits superpositions of configurations of flashes that then are somehow reduced to one configuration through wave-function collapse.

Furthermore, the rGRWf theory makes it possible to attribute a unique wave-function to the configuration of flashes on each particular spatial hypersurface given that hypersurface and all the flashes that occur earlier than it, and given the wave-function specified relative to any earlier hypersurface. The transition from the wave-function of the configuration of flashes on one hypersurface to the wave-function of the configuration of flashes on any later hypersurface also is Lorentz-invariant in the sense that it can be defined independently of the particular way in which the intervening space-time is foliated.

Let us now ask how, starting from an initial configuration of flashes and the initial wave-function, the actual distribution of flashes in space-time comes into being, and how the violation of locality established by Bell's theorem is implemented in the temporal development of the actual distribution of flashes (the following argument is based on Gisin 2010 and 2011). Consider, for the sake of simplicity, the EPR experiment, e.g. two 2-level atoms, located at a large distance from each other controlled by Alice and Bob, respectively. Assume that the distance is large enough so that Alice can collect the result of her measurement before any signal propagating at the speed of light can provide any information on Bob's choice of which measurement to perform; and *vice versa*, Bob's measurement outcome is separated by a spacelike interval from Alice's choice of measurement.

First, let us look at the situation from a reference frame in which Alice performs her measurement first. Denote by *x* her choice of which measurement to perform and by *a* the result she collects. The probability of her outcome can easily be computed and the process be simulated on a classical computer. For the purpose of such a simulation one needs a random number generator; one may use a quantum random number generator, but this is not necessary. Actually, one could merely fetch some numbers from a file that contains random numbers produced and saved to a hard disk a long time ago. Hence, Alice's probabilities can be thought of and simulated as if they were mere epistemic probabilities, that is as if the actual results were determined by some classical variable (the numbers stored on the computer's hard disk). Let us emphasize that all classical probabilities (i.e. those satisfying Kolmogorov's axioms) can be thought of and simulated in such a way. Denoting $\lambda$ the classical variables stored on the hard disk, Alice's result $a$ is a function of her measurement settings $x$ and of $\lambda$: $a=F_{AB}(x, \lambda)$, where $F_{AB}$ reminds us that this is the "first measurement" in the time order A-B.

Next, consider Bob's measurement in the same reference frame and denote it by $y$. His result, denoted by $b$, is also probabilistic, but, since he is second to measure, his result may depend also on Alice's measurement setting and outcome. In order to simulate Bob's outcome one may use the same random numbers fetched from the hard disk: $b=S_{AB}(x, y, a, \lambda)$, where $S_{AB}$ stands for "second" in the A-B time order. Note that one could think of $\lambda$ as a nonlocal variable. More precisely, one can conceive $\lambda$ as the physical state in the whole of the past light-cones of both $a$ and $b$, that is, as ranging over the entire distribution of flashes in both past light-cones, except for the measurement choices $x$ and $y$ that must remain free; however, for present purposes, it is sufficient to treat $\lambda$ as a variable used to simulate such experiments on one (local) classical computer.

That the above sketched simulation works, reproducing all probabilities and correlations as predicted by the quantum formalism, should be clear. Indeed, this is standard quantum mechanics and, again, this is how any stochastic process can be simulated. Let us now look at the experiment from another reference frame, one in which Bob is first to measure and Alice second. By symmetry it is clear that this situation can equally be simulated in the following manner: $b=F_{BA}(y, \lambda)$ and $a=S_{BA}(y, x, b, \lambda)$ with possibly different functions $F_{BA} \neq F_{AB}$, $S_{BA} \neq S_{AB}$. Note that the same file of $\lambda$'s can be used to simulate both cases, the one in which Alice is first and the one in which she is second. The variable $\lambda$ is arbitrarily large and the functions $F_{AB}$ and $F_{BA}$ may access different parts of $\lambda$. Averaging over $\lambda$, one recovers the probabilities for the set of all possible flashes, both in the case that Alice is first and in the case Bob is first.

In order to describe the occurrence of the actual distribution of flashes in a Lorentz-invariant way, there should be a file of appropriate $\lambda$'s, that is, of random variables of the appropriate size and structure, and 4 functions $F_{AB}$, $F_{BA}$, $S_{AB}$ and $S_{BA}$ such that the actual occurrence of the flashes does not depend on the reference frame, that is, does not depend on the arbitrary chronology, or, in other words, such that the actual distribution of the flashes comes about in a Lorentz-invariant manner:

$$a=F_{AB}(x, \lambda)=S_{BA}(y, x, b, \lambda) \qquad (1)$$
$$b=S_{AB}(x, y, a, \lambda)=F_{BA}(y, \lambda) \qquad (2)$$

for all $x$, $y$. However, no functions $F_{AB}$, $F_{BA}$, $S_{AB}$ and $S_{BA}$ satisfying (1) and (2) exist. Indeed, condition (2) implies that $S_{AB}$ is independent of $x$ and of $a$, which would turn $\lambda$ into a local variable (in Bell's sense, see Bell 2004, ch. 24, and Seevinck and Uffink 2011 for precisions). And it is well known that local variables cannot simulate all quantum correlations, in particular cannot simulate those violating a Bell inequality.

*Consequently, neither rGRWf, nor any other theory, can account for the occurrence of the Alice-flash and the occurrence of the Bob-flash in a Lorentz-invariant manner*. That is to say: even if one obtains the same outcomes and thus the same *final* distribution of the flashes in space-time, whatever reference frame one chooses, it is not possible to conceive the *coming*

*into being* of the flashes in a Lorentz-invariant manner. The reason is that the occurrence of some flashes depends on where in space-time other flashes occur: in one frame, Alice's outcome flash is independent of the flashes that constitute Bob's setting and outcome; in another frame, Alice's outcome flash depends on (or is influenced by) the flashes that constitute Bob's setting and outcome. The same goes for Bob's outcome flash. However, there is no Lorentz-invariant theory of these dependency relations or influences possible, as the argument above shows.

Tumulka concedes this point, but takes it to be irrelevant. He writes in his discussion of the EPR experiment:

> Who influences whom is frame dependent. There is no objective fact about who "really" influenced whom. There is no need for such a fact. The objective facts are where-when the flashes occur, and it is enough if a theory prescribes, as does rGRWf, their joint distribution in a Lorentz-invariant way. Whether nature chooses the space-time point $f_B$ first, and $f_A$ afterwards, or the other way around, does not seem like a meaningful question to me. (Tumulka 2007, p. 192; see also Tumulka 2009, pp. 209-210)

But if we want to know how the actual distribution of flashes comes into being, that question is meaningful: either the occurrence of the flash $f_A$ – Alice's outcome $a$ – depends only on Alice's measurement setting $x$ and λ, or it depends also on Bob's setting $y$ and Bob's outcome $b$. The same goes for the flash $f_B$. If a theory does not answer this question by taking either of these situations to be the real one (although, of course, we are not in the position to know which one of them is the real one), it does not provide for an ontology that takes the violation of Bell's condition of locality to be implemented in space-time, for it does not answer the question of whether or not the probabilities for a flash to occur at a given space-time point depend only on what there is in the past light-cone of that point or depend also on where flashes occur at spacelike separated intervals. In other words, if a theory endorses the existence of the flash $f_A$ (Alice's outcome $a$) and if a theory recognizes the existence of the Bell-type non-locality (or non-local influences, to take up Tumulka's expression) in space-time, then such a theory has to admit that there is a fact of the matter of whether or not the occurrence of the flash $f_A$ is dependent on (or influenced by) the occurrence of flashes at a space-like distance, such as the flash $f_B$ and the flashes of Bob's setting $y$. The same applies to any flash that occurs later than the initial configuration of flashes in any possible world of GRWf. In brief, if a theory does not admit the existence of such a fact, it is not a realism about the Bell-type non-local correlations, dependencies or influences in space-time.

However, there is no relativistic answer to these questions available: the mentioned facts presuppose a preferred foliation of space-time (which, however, does not have to be posed as additional space-time structure, but may be derivable from the universal wave-function – see Dürr et al. 2013b for such a proposal in the context of BM). By way of consequence, no relativistic theory of the actual occurrence of the flashes is possible, by contrast to the description of probabilities for joint flash distributions by means of a Lorentz-invariant law. In a nutshell, rGRWf is as Lorentz-invariant as possible, but it is not more Lorentz-invariant than possible, given the failure of Bell's locality condition.

One may have reservations about becoming, since there are serious arguments from relativity physics in favour of what is known as the block universe metaphysics. According to this metaphysics, everything that there is in space-time simply exists. To say the least, it would require much more argumentation than the one developed in this section to establish that quantum physics contradicts the block universe metaphysics, or the philosophical position of eternalism in general. What is important in the present context is that eternalism and the block universe metaphysics allow to formulate the question of how an initial configuration of matter actually develops in time (or, if they refused to admit that question, that refusal could then indeed be taken to count against that metaphysics). The common answer to that question is to insist on relativity physics and the block universe metaphysics making it possible to formulate a conception of local becoming, which is relativistic, namely becoming along a worldline (see e.g. Savitt 2001, Dieks 2006). The upshot of the argument of this section then is that GRWf does not admit a local becoming that is relativistic, for the becoming of some flashes depends on where other flashes occur at spacelike separated locations, and there is no relativistic answer to the question available of which flashes are subject to such dependency relations (or non-local influences) and which ones are not. In other words, the point is that the becoming that relativity physics is commonly taken to allow for, namely local becoming, cannot be obtained in a Lorentz-invariant manner in GRWf. Consequently, if there is a fact of the matter how the initial configuration of flashes actually develops so that one of the possible distributions of flashes becomes realized, there is a fact of the matter on which particular foliation of space-time into spatial hypersurfaces that are ordered in time that development occurs.

Instead of accepting the conclusion that there is one privileged, albeit unknown foliation of space-time into spatial hypersurfaces that are ordered in time, one may envisage to maintain that what there is in nature depends on the choice of a hypersurface – so that different facts exist in nature relative to the choice of a particular foliation of space-time.[6] However, if one is willing to endorse such a relativism, any of the known proposals for a primitive ontology of QM can then easily be made relativistic. Thus, in sum, Tumulka's rGRWf theory is relativistic only if one considers an entire distribution of flashes in space-time as the ontology of the theory. There is no room for a relativistic becoming in this ontology.

## 4. *Does the quantum state exist in addition to the primitive ontology?*

The commitment to a privileged foliation of space-time into spatial hypersurfaces that are ordered in time and hence the conflict between GRWf and relativity physics arises if and only if one acknowledges that the flashes stand in relations which make the occurrence of some flashes dependent on where other flashes occur, whereby these other flashes may be separated from the given flash by a spacelike interval. That is to say, over and above the flashes being primitive stuff in space-time – a flash occurring at a space-time point merely signifies that the point is not empty, but occupied by stuff –, what is known as the quantum state is instantiated in space-time. The fact that the quantum state is entangled then implies that the mentioned dependency relations exist among some flashes in space-time, whereby these relations are dynamically relevant for the development of the distribution of the flashes.

However, it is not mandatory to adopt a realist attitude to the quantum state in primitive ontology approaches to quantum physics. These approaches are endowed with an ontology in the form of the primitive ontology. One can therefore take the primitive ontology to be the *full* ontology. Consider what Bell says when introducing the notion of local beables:

---

[6] See Fleming (1996), Myrvold (2002) and Fine (2005, ch. 8, § 10, pp. 298-307) for such proposals.

> One of the apparent non-localities of quantum mechanics is the instantaneous, over all space, 'collapse of the wave function' on 'measurement'. But this does not bother us if we do not grant beable status to the wave function. We can regard it simply as a convenient but inessential mathematical device for formulating correlations between experimental procedures and experimental results, i.e., between one set of beables and another. Then its odd behaviour is acceptable as the funny behaviour of the scalar potential of Maxwell's theory in Coulomb gauge. (Quoted from Bell 2004, p. 53)

Not granting beable status to the wave-function means in this context that there is no quantum state over and above the local beables, that is, the distribution of the primitive stuff in space-time (such as the flashes). Given that *entire* distribution, the universal wave-function, figuring in the GRW law, is part of the *description* of the distribution of the primitive stuff in space-time (the flashes) that achieves the best balance between simplicity and empirical content. In order to do so, the *wave-function* correlates spacelike separated flashes, but there is no relation of entanglement (or of dependence or of influence) among the flashes instantiated in space-time over and above the local beables. In short, the world is the mosaic of individual flashes. These are simply there; their occurrence does not depend on anything.

Consequently, if one endorses Humeanism about laws, one can set out an ontology that is committed only to the distribution of the primitive stuff in the whole of space-time (see the references given in footnote 4 above). If that stuff are flashes, then that distribution is Lorentz-invariant. The availability of such a Humean ontology of quantum physics shows that quantum entanglement and Bell's theorem as such do not commit us to a non-Humean ontology that acknowledges modal facts or properties (for a prominent claim to the contrary, see e.g. Maudlin 2007a, pp. 51-64). However, note the price that one has to pay for this ontology: one has to forgo an account of the temporal development of the actual distribution of the flashes, and one can adopt realism only with respect to the primitive ontology, but not with respect to the quantum state.

By contrast, if one accords ontological significance to the wave-function, admitting the quantum state in the ontology and subscribes to a primitive ontology of matter distributed in space-time (such as flashes), then one is committed to recognizing physical relations existing among the elements of the primitive ontology (such as the flashes) over and above their mere location or occurrence at space-time points, namely relations of entanglement (dependence or influence) connecting certain such elements. In this case, there is no prospect of developing a Lorentz-invariant GRW flash ontology, since these relations connect in any case spacelike separated flashes; which flashes are linked by such relations is relevant for the dynamics of the temporal development of the distribution of the flashes, fixing whether (and if so how) the occurrence of each flash depends on where in space-time other flashes occur. By way of consequence, this view does not have to limit itself to endorsing only the whole distribution of flashes in space-time, but it has the means to include an account of the coming into being of the actual distribution of the flashes.

Consider dispositionalism as one possibility of an ontology that admits the quantum state in addition to the primitive ontology: on this view, the initial configuration of flashes instantiates a dispositional property, more precisely a propensity, which manifests itself in the occurrence of further flashes. The wave-function is not merely a tool for an economical description of the distribution of flashes in space-time as a whole, but it represents a property that the flashes have over and above being the primitive stuff in space-time, namely a property or propensity instantiated by a configuration of flashes that influences the temporal development of that configuration.

Moreover, not only the initial configuration of flashes instantiates that property, but all subsequent configurations also do so. These configurations involve in any case spacelike separated flashes. These spacelike separated flashes are related or connected in that they instantiate a holistic and dispositional property or propensity that manifests itself in the occurrence of further flashes, which themselves instantiate such a holistic and dispositional property or propensity, etc. Consequently, this ontology is committed to there being a fact of the matter which spacelike separated flashes are thus connected or related, because there is a fact of the matter which configurations of flashes instantiate that property. That fact is represented by the collapsed wave-function of these flashes. Accordingly, there is a fact of the matter in which temporal order the flashes occur in space, since certain configurations of spacelike separated flashes instantiate a dynamical property or propensity that influences the further temporal development of the distribution of the flashes, whereas others do not do so. On dispositionalism, in brief, not only the initial wave-function, but also the temporal development of that wave-function in the form of collapses on certain hypersurfaces has an ontological significance, referring to a property instantiated in space and time.

An analogous conclusion obtains if one endorses primitivism about laws and takes the laws to govern the temporal development of the objects to which they apply: on this view, any possible world instantiates the fact that a certain law holds in that world. If the law is the GRW law and if flashes are the primitive ontology, then the quantum state is the fact that the GRW law is instantiated by the flashes. On this view, again, the instantiation of that fact concerns not only the initial and the final, entire distribution of the flashes in space-time, but it influences the temporal development of that distribution, inducing a certain temporal development; in doing so, it connects certain configurations of spacelike separated flashes at a time. Consequently, which configurations of spacelike separated flashes are thus connected influences the further temporal development of the distribution of the flashes so that, again, a certain foliation of space-time into spatial hypersurfaces that are ordered in time is singled out.

In sum, thus, there is a relativistic GRWf theory only on the following two conditions: (a) one limits oneself to considering whole possible histories or distributions of flashes in space-time, and one renounces an account of the temporal development of the actual distribution of the flashes in space-time; (b) one subscribes to Humeanism, recognizing only the primitive stuff distributed in space-time, but no relations of entanglement (no quantum state) existing among the elements of that stuff in space-time.

## 5.   *How to account for interactions?*

The GRWf ontology is sparse in that there only are flashes occasionally occurring at space-time points. Nonetheless, it seems to be able to account for localized macroscopic objects including measurement devices: macroscopic objects are, as Bell put it, galaxies of flashes (Bell 2004, p. 205). The flash ontology is not an ontology of particles, but an ontology of events. However, this is not a problem in itself. There are serious arguments from relativity physics for four-dimensionalism, that is, the view that what there is in the world are four-dimensional entities such as events instead of three-dimensional ones changing in time such as particles; on this view, macroscopic objects are sequences of events that fulfil certain similarity criteria (see e.g. Sider 2001 and Balashov 2010 for contemporary four-dimensionalism). The flash ontology abandons the idea of these sequences being continuous; there is empty space between the events on the flash ontology. Nonetheless, the flash ontology allows for space-times in which there are enough flashes to constitute sequences of events that make up the macroscopic objects with which we are familiar. Maudlin (2011, pp. 257-258), however, points out that the flash ontology implies the radical falsity of our standard conception of small classical objects such as DNA strands, although he recognizes the empirical adequacy of this ontology. If a DNA strand consists of about $10^9$ atoms, its wave-function undergoes a GRW hit roughly once a day, so that there is a configuration of flashes constituting a DNA strand in space-time only once a day, instead of there being a continuous object, constituted by a continuous sequence of events on four-dimensionalism.

Against this background, another, even more serious problem can be raised for the flash ontology. Consider the question of what a measuring apparatus interacts with when it is supposed to measure a quantum system with mass such as an electron that has been prepared at a source and which the apparatus is designed to measure. For illustration, consider again the EPR experiment with a pair of electrons prepared at the source. Let us take for granted that the flash ontology can account for macroscopic objects such as measuring apparatuses in terms of "galaxies of flashes". But on the flash ontology, there is nothing between the source of the experiment and the measurement apparatus. In other words, there is nothing with which the apparatus could interact: there is no particle that enters it, no mass density and no field that gets in touch with it either. If the wave-function is a field, then, as mentioned in the second section, it can only be a field on configuration space; it does not exist in physical space-time. Consequently, whatever the ontological status of the wave-function may be, it cannot mediate the interaction among the flashes in space-time (as a classical field could do). There is only a pair of flashes in the past light-cone of the measuring apparatus, namely the two electron-flashes at the source of the experiment. That pair of flashes can be considered as having the propensity to bring about the occurrence of a further pair of flashes, and that propensity is supposed to be triggered by the measuring device; but that propensity is not a field that stretches out in space-time so that there is some physical entity or other with which the apparatus could interact.

On GRW quantum mechanics, the quantum system that is to be measured is supposed to be coupled with the huge configuration of quantum systems that make up the measuring apparatus, thereby to become entangled with that huge configuration, and that entanglement rapidly vanishes, since that huge configuration is immediately subject to a GRW hit. But this account does not make sense on the flash ontology: there is nothing with which the measuring apparatus could interact or which could be coupled to it (unless one were to stipulate that it directly and retroactively interacts with the flash or configuration of flashes in its past light-cone). Even if one attributes to the flashes the propensity to bring about further flashes, the question remains unanswered how to understand the triggering of this propensity through interactions, such as in measurement or other suitable interactions that provoke a GRW hit.

This consideration justifies drawing at least the following two conclusions: (a) the concern about becoming raised in the third section goes deeper than just touching the attempt to set out a relativistic flash ontology; even in the non-relativistic flash ontology, it remains unclear how the occurrence of flashes can be triggered through interactions. (b) As mentioned in the third section, Tumulka's rGRWf theory does not include an account of interactions; however, even before it comes to a formal account of interactions, the distribution of stuff in space-time is by far too sparse in GRWf in order to be able to even conceive the interaction of a measuring apparatus with a quantum system.

In conclusion, one may go for an event ontology instead of a particle ontology. But the flash ontology is too sparse an ontology: since it does not provide for anything like continuous sequences of events, it does not have the means at its disposal to account for interactions that are supposed to trigger the occurrence of further flashes (such as e.g. measurements). In the end, therefore, it seems that the flash ontology hardly is a convincing answer to the question of what quantum mechanics tells us about what there is in space-time.